\documentclass[11pt]{article}
\usepackage[centertags]{amsmath}
\usepackage{amsfonts}
\usepackage{amssymb}
\usepackage{amsthm}
\usepackage{newlfont}
\begin{document}
\setlength{\textwidth}{130mm} \setlength{\textheight}{194mm}

\title{Massive Particle Model with Spin from a Hybrid
(spacetime-twistorial) Phase Space Geometry and Its Quantization}

\author{Sergey  Fedoruk\thanks{BLTP JINR,
Dubna, Russia  \tt{e-mail:~fedoruk@theor.jinr.ru}} ,$\;$ Andrzej Frydryszak\thanks{ITP
University of Wroc{\l}aw, Poland \tt{e-mail:~amfry@ift.uni.wroc.pl}} ,$\;$ Jerzy
Lukierski\thanks{ITP University of Wroc{\l}aw, Poland \tt{e-mail:~lukier@ift.uni.wroc.pl}}
$\,$ and$\;$\\ \underline{C\`{e}sar Miquel-Espanya}\thanks{Dept. de F\'{\i}sica Te\`{o}rica
and IFIC, Val\`{e}ncia, Spain \& C.N.~YITP, Stony Brook, USA
\tt{e-mail:~Cesar.Miquel@ific.uv.es}} }
\date{} \maketitle

\begin{abstract}
We extend the Shirafuji model for massless particles with primary spacetime coordinates and
composite four-momenta to a model for massive particles with spin and electric charge. The
primary variables in the model are the spacetime four-vector, four scalars describing spin
and charge degrees of freedom as well as a pair of Weyl spinors. The geometric description
proposed in this paper provides an intermediate step between the free purely twistorial
model in two-twistor space in which both spacetime and four-momenta vectors are composite,
and the standard particle model, where both spacetime and four-momenta vectors are
elementary. We quantize the model and find explicitly the first-quantized wavefunctions
describing relativistic particles with mass, spin and electric charge. The spacetime
coordinates in the model are not commutative; this leads to a wavefunction that depends
only on one covariant projection of the spacetime four-vector defining plane wave solutions.
\end{abstract}

\section{Introduction}

There are three known equivalent ways of describing massless relativistic particles:
\begin{enumerate}
\item \textit{Purely twistorial description} - with primary twistor variables
and composite both spacetime and four-momenta \cite{Hu79};

\item \textit{Mixed twistorial-spacetime (Shirafuji-like) description} - with primary
spacetime coordinates and composite four-momenta \cite{Sh83}; and

\item \textit{Standard geometric description} - with primary
relativistic phase space variables (spacetime coordinates and four-momenta).
\end{enumerate}

The extension of these three geometric levels to the two-twistor sector has been presented
recently \cite{BeAzLuMi04,BeLuMi04,AzFrLuMi} in terms of the corresponding Liouville
one-forms. If we introduce two twistors ($i=1,2$; $A=1,\dots,4$)\footnote{The indices
$i=1,2$ describe an internal $SU(2)$ symmetry. The complex conjugation implies the change
from covariant (lower) indices to contravariant (upper) indices.}
\begin{equation*}
Z_{A i}= (\omega^\alpha_{\phantom{\alpha}i},\bar\pi_{{\dot\alpha} i})\, ,
\end{equation*}
the free Liouville one-form corresponding to the two-twistor case is the following
\begin{equation*}
\Theta_2=\frac{i}{2}\left(\omega^\alpha_{\phantom{\alpha}i}d\pi_{\alpha i}+
\bar\pi_{\dot{\alpha}i}d\bar\omega^{{\dot\alpha} i}-h.c.\right)\, .
\end{equation*}
After using the two-twistor generalization of the Penrose incidence relation and the
realization of the momentum $P_{\alpha\dot\alpha}$,
\begin{equation*}
P_{\alpha{\dot\beta}}=\pi_{\alpha}^{i}\bar\pi_{{\dot\beta} i}\ ,\label{Padb-2twistor}\qquad
\omega^\alpha_{\phantom{\alpha}i}= iz^{{\alpha{\dot\beta}}}\bar\pi_{{\dot\beta} i}\ ,\qquad
\text{where}\quad z^{{\alpha{\dot\beta}}}=x^{\alpha{\dot\beta}} + i y^{\alpha{\dot\beta}}\,
,
\end{equation*}
one obtains $\Theta'_2=\pi_\alpha^{\phantom{\alpha}i}\bar\pi_{{\dot\beta}
i}dx^{\alpha{\dot\beta}}+
iy^{\alpha{\dot\beta}}(\pi_\alpha^{\phantom{\alpha}i}d\bar\pi_{{\dot\beta} i}-
\bar\pi_{{\dot\beta} i} d\pi_\alpha^{\phantom{\alpha}i})$.

We introduce new variables $s^i_{\ j}=-2y^{\alpha{\dot\beta}}
\pi_{\alpha}^{i}\bar\pi_{\dot\beta j}=\overline{(s^j_{\ i})}$, and define $f$ and
${\overline{f}}$ satisfying
\begin{eqnarray*}
\bar\pi^{{\dot\alpha} i}\bar\pi_{\dot\alpha}^{\phantom{\alpha}j}=-\epsilon^{ij}f\quad
,\label{pipi1}&\qquad& \pi^\alpha_{\phantom{\alpha}i}\pi_{\alpha j}=-\epsilon_{ij}\bar
f\quad
,\\
\bar\pi_{{\dot\alpha}
i}\bar\pi_{{\dot\beta}}^{\phantom{\alpha}i}=\epsilon_{{\dot\alpha}{\dot\beta}}f\quad
,&\qquad& \pi_{\alpha i}\pi_\beta^{\phantom{\beta} i} =\epsilon_{\alpha\beta}\bar f\,
.\label{pipi4}
\end{eqnarray*}
Now, we can write
\begin{equation}\label{LiouvillePrime2New}
\Theta'_2= \pi_\alpha^{\phantom{\alpha}i}\bar\pi_{{\dot\alpha} i}
dx^{\alpha\dot\alpha}+\frac{i}{2}s_i^{\ j}\left[\frac{1}{\bar
f}\pi^\alpha_{\phantom{\alpha}k}d\pi_{\alpha j}\epsilon^{ki}
+\frac{1}{f}{\overline{\pi}}^{{\dot\alpha}
i}d\bar\pi_{{\dot\alpha}}^{\phantom{\alpha}k}\epsilon_{kj}\right]\, .
\end{equation}

Formula (\ref{LiouvillePrime2New}) determines the two-twistor generalization of the
Shirafuji action \cite{Sh83}. We see that the primary or, equivalently, elementary
variables are now the following ones
\begin{eqnarray*}
N=1 &\Rightarrow& N=2\nonumber\\
x^{\alpha{\dot\alpha}}, \pi_\alpha,\bar\pi_{\dot\alpha}&\Rightarrow&
x^{\alpha{\dot\alpha}},\pi_{\alpha i},\bar\pi_{\dot\alpha}^{\phantom{\alpha}i},s_i^{\ j}\,
.
\end{eqnarray*}
The particle model described by the Liouville one-form (\ref{LiouvillePrime2New}) provides
a framework to describe the mass, spin and electric charge but does not specify their
values. We shall introduce further their numerical values by postulating suitable physical
constraints. One concludes that the quantum-mechanical solution of the model
(\ref{LiouvillePrime2New}) may describe infinite-dimensional higher spin and electric
charge multiplets, linked with the field-theoretic formulation of higher spin theories (see
e.g. \cite{Vas,Sez,Sor}).

\section{The classical model - analysis of constraints in phase space}\label{Sect.An.Constaints}

\subsection{Action, conservation laws and physical constraints}
The dynamics of a massive spinning particle is described by its trajectory in the
generalized coordinate space
\begin{equation}\label{coord}
Q_L (\tau) =\left( x^\mu(\tau), \pi_{\alpha k}(\tau), \bar\pi_{\dot\alpha}^k(\tau),
s_k{}^j(\tau)\right)\, ,
\end{equation}
where $\pi_{\alpha k}$, $\bar\pi_{\dot\alpha}^k=\overline{(\pi_{\alpha k})}$ ($k,j=1,2$)
are two pairs of commuting Weyl spinors and the four quantities $s_k{}^j$, satisfying the
condition $s_k{}^j=\overline{(s_j{}^k)}$, are Lorentz scalars. The action derived from
(\ref{LiouvillePrime2New}) has the following form $(a=1,\dots,4)$
\begin{equation}\label{act}
S=\int d\tau\,{\cal L} = \int
d\tau\,\left[\pi_\alpha^{\phantom{\alpha}i}\bar\pi_{{\dot\alpha} i}\dot
x^{\alpha\dot\alpha} + {\frac{i}{2}}s_k{}^j \left({\frac{1}{\bar f}}\pi^{\alpha
k}\dot\pi_{\alpha j}
+{\frac{1}{f}}\bar\pi^{\dot\alpha}_j\dot{\bar\pi}_{\dot\alpha}^k\right) +\lambda^a
T_a\right]\, ,
\end{equation}
where the $T_a$ are four algebraic constraints on the coordinates (\ref{coord}) to be
specified later and $\lambda^a$ are their Lagrange multipliers.

{}From the Lagrangian (\ref{act}) we obtain the canonical momenta in the standard way
($\mathcal{P}^L=\frac{\partial \mathcal{L}}{\partial \dot{\mathcal{Q}}_L}$). This leads to the
following 16 primary constraints
\begin{equation}\label{c:p-x}
P^{\alpha\dot\alpha} -\pi^{\alpha k}\bar\pi^{\dot\alpha}_k\approx 0\, ,\quad
P_{(s)}{}^k{}_j \approx 0\, ,
\end{equation}
\begin{equation}\label{c:p-pi}
P^{\alpha j} -\frac{i}{2\bar f}\pi^{\alpha k}s_k{}^j \approx 0\quad , \qquad \bar
P^{\dot\alpha}_j - \frac{i}{2f} s_j{}^k \bar\pi^{\dot\alpha}_k \approx 0\, ,
\end{equation}

We present now the form of the four algebraic constraints\footnote{The justification
of the form of the constraints $T_a$ can be obtained by considering the symmetries of the
action (\ref{act}). It appears that the choice (\ref{mass})-(\ref{Q}) and the
interpretation of $s_i$ as covariant spin projection is related with the formulae for the
corresponding Noether charges.} $T_a$

\begin{eqnarray}
T_1\,: & \qquad &T \equiv 4f\bar f-m^2\approx 0 \, , \label{mass}\\
T_2\,: & \qquad &S \equiv {\mathbf{s}}^2-s(s+1)\approx 0 \, , \label{S}\\
T_3\,: & \qquad &S_3 \equiv s_3-m_3\approx 0 \, , \label{S-3}\\
T_4\,: & \qquad &Q \equiv s_0-q\approx 0 \, . \label{Q}
\end{eqnarray}
The real quantities ${\mathbf{s}}=(s_r)= (s_1, s_2, s_3)$ and $s_0$ present in
(\ref{S})-(\ref{Q}) are defined in terms of the Lagrangian variables $s_k^j$ as follows
\begin{equation*}\label{lagva2.18}
s_0={\frac{1}{2}}s_k{}^k\, ,\qquad s_r={\frac{1}{2}}s_k{}^j (\sigma_r)_j{}^k \, ,\qquad
r=1,2,3\, ,
\end{equation*}
where $(\sigma_r)_j{}^k$ are the Pauli matrices.

The constraint (\ref{mass}) defines the mass $m$ of the particle because using it we obtain
that ($P_\mu\equiv \sigma_{\mu\alpha\dot\alpha}P^{\alpha\dot\alpha}$) $ P_\mu P^\mu= m^2$.
The constraints (\ref{S}) and (\ref{S-3}) are introduced in the action (\ref{act}) in order
to obtain a definite spin $s$ and the covariant spin projection $s_3$ whereas the
constraint (\ref{Q}) defines the $U(1)$ charge $q$ of the particle. The case of a massive spinless particle has recently been
described in terms of a single twistor by using a modified
twistor-phase space transform inspired by two-time
physics techniques \cite{BarsPicon}.

In the subsection \ref{Sect.TimeEvolution} we shall see, from the time preservation of the
constraints, that secondary constraints do not appear in our model. Thus, the full
set of constraints is given by the physical constraints (\ref{mass})-(\ref{Q}) and by the
primary ones (\ref{c:p-x})-(\ref{c:p-pi}).

\subsection{Analysis of the primary constraints}

If we transform the twelve constraints (\ref{c:p-x}), (\ref{c:p-pi}) to equivalent
Lorentz-invariant expressions by contracting them with the spinors $\pi_{\alpha k}$ and
$\bar \pi_{\dot\alpha}^k$ and combine the results, the discussion of the constraints is
simplified and their splitting into first and second class is clearer. The eight
expressions (\ref{c:p-pi}) take the form
\begin{equation*}\label{D,B}
D_k{}^j \equiv {\cal D}_k{}^j +s_k{}^j \approx 0\, , \qquad B_k{}^j \equiv {\cal B}_k{}^j
\approx 0\, ,
\end{equation*}
where ${\cal D}_k{}^j \equiv i(\pi_{\alpha k}P^{\alpha j} - \bar P^{\dot\alpha}_k \bar
\pi_{\dot\alpha}^j) \, , \qquad {\cal B}_k{}^j \equiv i(\pi_{\alpha k}P^{\alpha j} + \bar
P^{\dot\alpha}_k \bar \pi_{\dot\alpha}^j)$.

The first four constraints of (\ref{c:p-x}), after contraction with spinors and using the
mass shell constraint, take the form
\begin{equation*}\label{Px}
C_k{}^l \equiv {\cal P}_k{}^l +m^2\delta_k{}^l \approx 0\quad, \qquad
\text{where}\quad{\cal P}_k{}^l \equiv 4 \pi_{\alpha
k}P^{\alpha\dot\beta}\bar\pi_{\dot\beta}^{\phantom{\beta}l}\, .
\end{equation*}

We transform the new set of 16 primary constraints, in order to get $SU(2)$ vector and
scalar quantities, in the following way
\begin{equation*}
A_r=\frac{1}{2}A_i^j(\sigma_r)_j^i\quad,\qquad A_0=\frac{1}{2}A_i^i\quad,\qquad
A=\{B,C,D,P_{(s)}\}\quad,
\end{equation*}
where $(\sigma_r)_j{}^k$, $r=1,2,3$ are the Pauli matrices. The set of primary
constraints takes the following form
\begin{eqnarray}
R_r \equiv P_{(s)}{}_r \approx 0\, , & \qquad &R_0 \equiv P_{(s)}{}_0
\approx 0\, , \label{const-Ps}\\
D_r \equiv {\cal D}_r +s_r \approx 0\, , & \qquad &D_0 \equiv {\cal
D}_0+s_0 \approx 0\, , \label{const-D}\\
B_r \equiv {\cal B}_r \approx 0\, , & \qquad &B_0 \equiv {\cal B}_0
\approx 0\, , \label{const-B}\\
C_r \equiv {\cal P}_r \approx 0 \, , & \qquad &C_0 \equiv {\cal P}_0 +m^2 \approx 0\, .
\label{const-Px}
\end{eqnarray}
Thus, our full set of constraints is described now by the four physical constraints
(\ref{mass})-(\ref{Q}) and by the sixteen primary constraints
(\ref{const-Ps})-(\ref{const-Px}).

We present now the canonical Poisson brackets (PB) of the coordinates $\mathcal{Q}_L$ and their
momenta $\mathcal{P}_L$
\begin{eqnarray*}
\{ x^\mu, P_\nu\}=\delta^\mu_\nu \, ,&\qquad &\{ s_k{}^j,
P_{(s)}{}^n{}_l\}=\delta_k^n\delta^j_l
\, ,\label{CBxP}\\
\{ \pi_{\alpha k}, P^{\beta j}\}=\delta_\alpha^\beta \delta_k^j \, ,&\qquad& \{
{\bar\pi}_{\dot\alpha}^k, \bar P^{\dot\beta}_j\}= \delta_{\dot\alpha}^{\dot\beta}\delta^k_j \, ,\\
\{ s_0, P_{(s)}{}_0\}=\frac{1}{2}\, ,&\qquad&\{ s_r, P_{(s)}{}_q\}=\frac{1}{2}\delta_{rq}\,
.\label{CBsrPs}
\end{eqnarray*}

Evaluating the corresponding Poisson brackets we see that the three quantities
$\mathcal{D}_r$ are the generators of $SO(3)$ and the three quantities $\mathcal {B}_r$
extend the $\mathcal{SO}(3)$ algebra to the Lorentz symmetry $SO(3,1)\simeq
sl(2;\mathbb{C})$. We shall call the $\mathcal{D}_r$, $\mathcal{B}_r$ internal symmetry generators.

We can also conclude that the quantities ${\cal P}_0$, ${\cal P}_r$ extend the internal
Lorentz generators $({\cal D}_r, {\cal B}_r)$ to an internal Poincar\'{e} algebra. The
complete set of non-vanishing Poisson brackets between all twenty constraints can be found
in \cite{Fed06}.

\subsection{Time evolution of constraints and their split into first and second
class ones}\label{Sect.TimeEvolution}

The action (\ref{act}) is invariant under an arbitrary rescaling on the world line
$\tau\rightarrow \tau'=\tau'(\tau)$ and the canonical Hamiltonian vanishes
$\mathcal{H}=\mathcal{P}_L \dot{{Q}}_L -\mathcal{L}=0\,$. The total Hamiltonian is given,
therefore, by a linear combination of all the constraints,
\begin{eqnarray*}
{\cal H}^C & = & \lambda^{(D)}_r D_r+ \lambda^{(D)}_0 D_0+ \lambda^{(B)}_r B_r+
\lambda^{(B)}_0 B_0+ \lambda^{(C)}_r C_r+
\lambda^{(C)}_0 C_0+ \nonumber\\
&& +\lambda^{(R)}_r R_r+ \lambda^{(R)}_0 R_0+ \lambda^{(T)} T+ \lambda^{(S)} S+
\lambda^{(S_3)} S_3+ \lambda^{(Q)} Q\, . \label{H}
\end{eqnarray*}

Imposing the preservation of all the constraints in time we find that four out of the twenty
Lagrange multipliers above are not determined. The four first class constraints associated with
them are
\begin{eqnarray}\label{cF}
{\cal F} = C_0 + {\frac{1}{2}}T \simeq 0 \, &,& {\cal S} = S- 2s_r
D_r \simeq 0 \, ,\\
{\cal S}_3 = S_3-D_3- 2\epsilon_{3rq}s_q R_r \simeq 0 \, &,& {\cal Q} = Q-D_0 \simeq
0\,.\label{cQ}
\end{eqnarray}
The other 16 constraints can be presented as eight pairs of canonically conjugated second
class constraints $ (D_r\Leftrightarrow R_r \, ,\ D_0\Leftrightarrow R_0\,,\ B_r
\Leftrightarrow C_r\,,\ B_0 \Leftrightarrow T)$.

The subset of constraints $D_r$, $B_r$ does not close under the PB operation. To solve this
we introduce the following linear combination of constraints
\begin{eqnarray*}\label{D-pr}
D_r{}^\prime &\equiv& D_r-\epsilon_{rpq}s_pR_q={\cal D}_r +s_r
-\epsilon_{rpq}s_p{\cal P}_{(s)}{}_q \approx 0\, , \\
B_r{}^\prime &\equiv& B_r+ {\frac{i}{2m^2}}\epsilon_{rpq}s_p C_q={\cal B}_r+
{\frac{i}{2m^2}}\epsilon_{rpq}s_p {\cal P}_q \approx 0\, .
\end{eqnarray*}
Now, their Poisson brackets vanish on the surface of the constraints.

\section{Solving the second class
constraints}\label{Sect.Sol.2nd.Class.Constr}

Due to the resolution form\footnote{A pair of constraints $A\approx 0,B\approx 0$ have the
{\it resolution form} in the phase space $(x_i,p_i)$ $i=1,\dots,N$ if they have the form given by
the following formulae:
\begin{equation*}
A=x_1-f(x_r,p_r)\approx 0\qquad B=p_1\approx 0\quad (r=2,3,\dots,N)
\end{equation*}
This form of the constraints was considered by Dirac \cite{Dirac}. In such a case the Dirac
brackets are identical with the canonical PB.} of the first four pairs of second class
constraints we can exclude \cite{Fed06} the variables $s_r$, $s_0$ and their momenta
$P_{(s)}{}_r$, $P_{(s)}{}_0$ by putting instead
\begin{equation*}\label{s}
s_r = - {\cal D}_r \, ,\qquad s_0=- D_0 \, ,\qquad P_{(s)}{}_r = 0 \, ,\qquad P_{(s)}{}_0
=0 \, .
\end{equation*}

In order to put the next three pairs of second class constraints in the resolution form we
perform a canonical transformation\footnote{The generating function of this canonical
transformation has the form
\begin{eqnarray}
&F(P^\mu, \pi_{\alpha k}, \bar\pi_{\dot\beta}^k; \widetilde{x}_0, \widetilde{x}_r, {\cal
P}^{\alpha k}, \bar{\cal
P}^{\dot\alpha}_k)=&\\
&= -[\pi_{\alpha k}\sigma_\mu^{\alpha\dot\beta}\bar\pi_{\dot\beta}^k P^\mu]\widetilde{x}_0
+ [(\tau_r)_j{}^k\pi_{\alpha k}\sigma_\mu^{\alpha\dot\beta}\bar\pi_{\dot\beta}^j P^\mu]
\widetilde{x}_r + \pi_{\alpha k}{\cal P}^{\alpha k} +\bar\pi_{\dot\alpha}^k \bar{\cal
P}^{\dot\alpha}_k \, . &
\end{eqnarray}}
\begin{equation*}
(x^\mu\, ; P_\mu)\, , \ (\pi_{\alpha k}\, ; P^{\alpha k})\, , \ (\bar\pi_{\dot\alpha}^k\, ;
\bar P^{\dot\alpha}_k)\ \Leftrightarrow \ (\widetilde{x}_0, \widetilde{x}_r\, ; {\cal P}_0
, {\cal P}_r)\, , \ (\pi^\prime_{\alpha k}\, ; {\cal P}^{\alpha k})\, , \ (\bar\pi^{\prime
k}_{\dot\alpha}\,; \bar{\cal P}^{\dot\alpha}_k) \, .
\end{equation*}

Now, we can exclude the variables $\widetilde{x}_r$ and $\mathcal{P}_r$ by setting
\begin{equation*}
\widetilde{x}_r =-\frac{i}{{\cal P}_0}{\cal B}_r\, , \qquad {\cal P}_r =0\, .
\end{equation*}

The only two remaining second class constraints have the form
\begin{eqnarray}
T=4f{\overline{f}} -m^2=0\, ,&&\label{T.Constraint}
B_0=\mathcal{B}_0-i\widetilde{x}_0\mathcal{P}_0=0\, .
\end{eqnarray}
Subsequently, we introduce the Dirac brackets (DB) as follows
\begin{equation*}\label{DB2}
\{y, y'\}_{D}=\{y,y'\}+\{y, B_0\}{\frac{i}{2(T+m^2)}}\{T, y'\} -\{y,
T\}{\frac{i}{2(T+m^2)}}\{B_0, y'\}\, ,
\end{equation*}

We observe that the first relation of (\ref{T.Constraint}) reduces one spinorial degree of
freedom, {\it i.e.} we are left with seven unconstrained spinorial coordinates.

\section{First quantization and solution
of the first class constraints}\label{Sect.Sol.1st.Class.Constr}

\subsection{First class constraints}

After taking into account all the sixteen second class constraints there remain  eighteen
phase space variables, namely,
\begin{equation*}\label{newvariables}
\widetilde{x}_0\, ,\,\, {\cal P}_0\quad ;\qquad \pi_{\alpha k}\, ,\,\, {\cal P}^{\alpha k};
\qquad \bar\pi^{k}_{\dot\alpha}\, ,\,\, \bar{\cal P}^{\dot\alpha}_k\quad ,
\end{equation*}
which are constrained by the two algebraic relations (\ref{T.Constraint}). After performing the
quantization of the canonical Dirac brackets $\{y,y'\}_D\rightarrow\frac{1}{i}[\hat y,\hat
y']$ (we put $\hbar=1$) one obtains the corresponding commutation relations, where we use
the `$qp$-ordering'.

The sixteen independent degrees of freedom described by the variables (\ref{newvariables})
are additionally restricted by the four first class constraints (\ref{cF})-(\ref{cQ}).
These can be written in the form
\begin{eqnarray}\label{1}
{\cal P}_0 +m^2 &\approx& 0\, ,\\
\label{2}
{\cal D}_r {\cal D}_r -s(s+1) &\approx& 0\, ,\\
\label{3}
{\cal D}_3 +m_3 &\approx& 0\, ,\\
\label{4}
{\cal D}_0 +q &\approx& 0\, ,
\end{eqnarray}
where the numerical values of $m,s,m_3$ and $q$ describe mass, spin, spin projection and an
internal Abelian (electric) charge.

\subsection{Covariant solution of the constraints}\label{Sect.Covar.Sol}

We take the Schr\"{o}dinger realization of the quantized variables (\ref{newvariables}) on
the commuting generalized coordinate space $(\widetilde{x}_0,\pi_{\alpha j},
{\overline{\pi}}_{\dot\alpha}^j)$. The generalized momenta
$(\mathcal{P}_0,\mathcal{P}^{\beta
j},\overline{\mathcal{P}}^{\dot\beta}_{\phantom{\beta}j})$ have the following differential
realizations:
\begin{equation*}\label{calP-0}
{\cal P}_0=-i\frac{\partial}{\partial \widetilde{x}_0}\, ,
\end{equation*}
\begin{equation*}\label{calP-a}
{\cal P}^{\beta j}=-i\frac{\partial}{\partial\pi_{\beta j}} +i\frac{f}{m^2}\pi^{\beta
j}\left(\pi_{\alpha k}\frac{\partial}{\partial\pi_{\alpha k}} + \bar \pi_{\dot\alpha}^k
\frac{\partial}{\partial\bar \pi_{\dot\alpha}^k} -2\widetilde{x}_0\frac{\partial}{\partial
\widetilde{x}_0}\right)\, ,
\end{equation*}\
\begin{equation*}\label{calP-b}
\bar {\cal P}^{\dot\beta}_j=-i\frac{\partial}{\partial\bar\pi^{j}_{\dot\beta}} -i\frac{\bar
f}{m^2}\bar\pi_{j}^{\dot\beta}\left(\pi_{\alpha k}\frac{\partial}{\partial\pi_{\alpha k}} +
\bar \pi_{\dot\alpha}^k \frac{\partial}{\partial\bar \pi_{\dot\alpha}^k}
-2\widetilde{x}_0\frac{\partial}{\partial \widetilde{x}_0}\right)\, .
\end{equation*}

The wavefunction has the following coordinate dependence
\begin{equation*}
\Psi(\widetilde{x}_0, \pi_{\alpha k}, \bar \pi_{\dot\alpha}^k)\,.
\end{equation*}
We shall solve the four wave equations consecutively:

\medskip
\textit{i) Mass shell constraint (\ref{1}).}

The general solution of the constraint (\ref{1}) is
\begin{equation}\label{wf-mass}
\Psi(\widetilde{x}_0, \pi_{\alpha k}, \bar \pi_{\dot\alpha}^k)= e^{-im^2 \widetilde{x}_0}\,
\Phi(\pi_{\alpha k}, \bar \pi_{\dot\alpha}^k)\, .
\end{equation}
One can show \cite{Fed06} that the invariant time coordinate
$\widetilde{\tau}=m^2\widetilde{x}_0=P^\mu x_\mu$. Therefore, the exponent in the
wavefunction (\ref{wf-mass}) has the standard form of a plane wave, where the four-momentum
$P_\mu$ is composite in terms of spinors.

\medskip
\textit{ii) Normalized spinors and electric charge.}

The eight (real) variables $\pi_{\alpha k}$, $\bar \pi_{\dot\alpha}^k$ define two variables
$f$, $\bar f$ as follows
\begin{equation*}\label{ff}
\pi^{\alpha k}\pi_{\alpha k}=2\bar f \, , \qquad \bar\pi_{\dot\alpha k}\bar\pi^{\dot\alpha
k} =2f
\end{equation*}
and the remaining six degrees of freedom can be described by the normalized spinors
$u_{\alpha i}= \left({{\frac{\bar f}{f}}}\right)^{-1/4}\pi_{\alpha i}\, , \quad \bar
u_{\dot\alpha}^i= \overline{(u_{\alpha i})}=\left({{\frac{\bar f}{f}}}\right)^{1/4}
\bar\pi_{\dot\alpha}^i\, .$

Due to the first constraint of (\ref{T.Constraint}), the modulus of $f$ is given by the mass
parameter $ |f|={{\frac{m}{2}}}$ and the variable $y\in S^1,\quad y\equiv {{\frac{\bar
f}{f}}}$, defines the phase of $f$ which will be eliminated by the constraint (\ref{4}).

Using the variables $y$, $u_{\alpha i}$, $\bar u_{\dot\alpha}^i$ the first class constraint
(\ref{4}) has the solution
\begin{equation*}\label{Phi-b}
\Phi_m(y, u_{\alpha k}, \bar u_{\dot\alpha}^k)= y^{-q/2} \tilde\Phi_m(u_{\alpha k}, \bar
u_{\dot\alpha}^k)\, .
\end{equation*}

\medskip
{\it iii) Spin content.}

Let us find now the solution of the constraints (\ref{2}), (\ref{3}) for the function
$\tilde\Phi(u_{\alpha k}, \bar u_{\dot\alpha}^k)$ using the polynomial expansion in spinor
variables
\begin{equation*}\label{Phi}
\tilde\Phi(u_{\alpha k}, \bar u_{\dot\alpha}^k)= \sum_{k,n=0}^{\infty} {\frac{1}{k!\,n!}}\,
u_{\alpha_1 i_1}\ldots u_{\alpha_k i_k} \bar u_{\dot\beta_1}^{j_1}\ldots\bar
u_{\dot\beta_n}^{j_n}\, \phi^{\alpha_1\ldots\alpha_k
\dot\beta_1\ldots\dot\beta_n}{}^{i_1\ldots i_k}_{j_1\ldots j_n}(P_\mu)\, .
\end{equation*}

The solution of Eq. (\ref{2}) is
\begin{equation*}\label{sol}
\tilde\Phi(u_{\alpha k}, \bar u_{\dot\alpha}^k)= \sum_{{k,n;\, k+n=2s}}
{\frac{1}{k!\,n!}}\, u_{\alpha_1 i_1}\ldots u_{\alpha_k i_k} \bar
u_{\dot\beta_1}^{j_1}\ldots\bar u_{\dot\beta_n}^{j_n}\, \phi^{\alpha_1\ldots\alpha_k
\dot\beta_1\ldots\dot\beta_n}{}^{i_1\ldots i_k}_{j_1\ldots j_n}(P_\mu)\, ,
\end{equation*}
where in this expansion only spinorial polynomials of order $2s$ ($k=k_1+k_2$, $n=n_1+n_2$)
are present, $k+n=2s\, ,$ where $k_i$ $(i=1,2)$ denotes the number of spinors $u^\alpha_i$, and $n_i$ $(i=1,2)$ the
number of spinors $\overline{u}_i^{\dot\alpha}$.

In order to describe covariant projection of the spin, given by the eigenvalue equation
(\ref{3}), we observe that
\begin{eqnarray*}
&(\mathcal{D}_3+m_3)(u_{\alpha_1i_1}\cdots
u_{\alpha_ki_k}\overline{u}^{j_1}_{{\dot\beta}_1}\cdots\overline{u}^{j_n}_{{\dot\beta}_n})=0\
,& \\
&\text{if} \qquad m_3=\frac{1}{2}(n_1-n_2-(k_1-k_2))\, .&
\end{eqnarray*}

A more detailed discussion of the procedure used to obtain the solutions, the link between
the variables $u^\alpha_i$ and the spinorial Lorentz harmonics can be found in \cite{Fed06}. Although we
are going to present here the example for the spin 1 wavefunctions the simpler
spin $\frac{1}{2}$ case can also be found in \cite{AzFrLuMi,Fed06}.

\section{Example: Spin $s=1$.}

In this case the field $\tilde\Phi(u_{\alpha k}, \bar u_{\dot\alpha}^k)$ is
\begin{equation}\label{sol-1}
\tilde\Phi(u_{\alpha k}, \bar u_{\dot\alpha}^k)= {\frac{1}{2}}\,
u_{\alpha i} u_{\beta j} \, \phi^{\alpha\beta}{}^{ij}+ u_{\alpha
i}\bar u_{\dot\beta}^{j}\, \phi^{\alpha \dot\beta}{}^{i}_{j}+
{\frac{1}{2}}\, \bar u_{\dot\alpha}^{i}\bar u_{\dot\beta}^{j}\,
\phi^{\dot\alpha\dot\beta}{}_{ij}\, .
\end{equation}
Inserting in this expression $u_{\alpha i
}={{\frac{2}{m}}}P_{\alpha\dot\alpha}\bar u^{\dot\alpha}_i$, $\bar
u_{\dot\alpha}^i=-{{\frac{2}{m}}}P_{\alpha\dot\alpha} u^{\alpha i}$ and comparing the result with (\ref{sol-1}) we
obtain the following equations
\begin{eqnarray}\label{Dir-1a}
&P_{\alpha\dot\alpha}\phi^{\dot\alpha}_{\beta}{}^{ij}+
{{\frac{m}{2}}} \phi_{\alpha\beta}{}^{ij}=0\, , \qquad
P^{\dot\alpha\alpha} \phi_{\alpha}^{\dot \beta}{}^{ij}+
{{\frac{m}{2}}}\phi^{\dot\alpha\dot\beta}{}^{ij}=0\, ,&\\
\label{Dir-1b}
&{\frac{1}{2}}\,(P_{\alpha\dot\alpha}\phi^{\dot\alpha
\dot\beta}{}^{ij}+  P^{\dot\beta\beta} \phi_{\alpha\beta}{}^{ij})+
{{\frac{m}{2}}}\phi_{\alpha}^{\dot\beta}{}^{ij}=0\, .&
\end{eqnarray}
The antisymmetric parts of equations (\ref{Dir-1a}) provide the
transversality condition for fields $\phi^{\alpha \dot\beta}{}_i^j$
\begin{equation}\label{trans}
P_{\alpha\dot\beta}\phi^{\alpha \dot\beta}{}^{ij}=0\, .
\end{equation}
Using $P_{\alpha\dot\beta}P^{\dot\beta\beta}=
{{\frac{1}{4}}}m^2\delta_\alpha^\beta$ we obtain further
\begin{equation}\label{Dir-1c}
P_{\alpha\dot\alpha}\phi^{\dot\alpha \dot\beta}{}^{ij}+
{{\frac{m}{2}}}\phi_{\alpha}^{\dot\beta}{}^{ij}=0\, , \qquad
P^{\dot\beta\beta} \phi_{\alpha\beta}{}^{ij}+
{{\frac{m}{2}}}\phi_{\alpha}^{\dot\beta}{}^{ij}=0\, .
\end{equation}

The equations (\ref{Dir-1a})-(\ref{Dir-1c}) are the Bargmann-Wigner
equations written in a two-spinor notation. One can pass to the
four-component Dirac spinor notation if one constructs from the
fields $\phi_{\alpha\beta}{}^{ij}$,
$\phi^{\dot\alpha\dot\beta}{}^{ij}$,
$\phi_{\alpha}{}^{\dot\beta}{}^{ij}$ and
$\phi^{\dot\beta}{}_{\alpha}{}{}^{ij}\equiv
\phi_{\alpha}{}^{\dot\beta}{}{}^{ij}$ the following Bargmann-Wigner
fields
$
\psi_{ab}{}^{ij}= \left(
\begin{array}{c}
\phi_{\alpha b}{}{}^{ij} \\
\phi^{\dot\alpha}{}_{b}{}^{ij} \\
\end{array}
\right)= \left(
\begin{array}{c}
\phi_{a\beta}{}^{ij} \\
\phi_{a}{}^{\dot\beta}{}^{ij} \\
\end{array}
\right)\, ,
$
with double Dirac indices $a,b=1,2,3,4$. Since
$\phi_{\alpha\beta}{}^{ij}=\phi_{\beta\alpha}{}^{ij}$,
$\phi^{\dot\alpha\dot\beta}{}^{ij}=\phi^{\dot\beta\dot\alpha}{}^{ij}$
the fields $\psi_{ab}{}^{ij}$ are symmetric,
$\psi_{ab}{}{}^{ij}=\psi_{ba}{}^{ij}$. Due to the equations
(\ref{Dir-1a})-(\ref{Dir-1c}) the fields $\psi_{ab}{}^{ij}$ satisfy the
Bargmann-Wigner-Dirac equation for massive spin 1 fields: $P^\mu \,\gamma_{\mu a}{}^{b}
\,\psi_{bc}{}^{ij}+m\psi_{ac}{}^{ij}=0\, .$

We obtain Proca fields if we define the fields
\begin{equation} \label{Proca}
A_\mu{}^{ij} =
\sigma_\mu{}^\alpha_{\dot\beta}\phi_\alpha{}^{\dot\beta}{}^{ij}\quad
,\qquad F_{\mu\nu}{}^{ij} = m(\sigma_{\mu\nu}{}^{\alpha}_{\beta}
\phi_{\alpha}^{\beta}{}^{ij}
+\bar\sigma_{\mu\nu}{}^{\dot\alpha}_{\dot\beta}
\phi_{\dot\alpha}^{\dot\beta}{}^{ij})\, .
\end{equation}
Inserting (\ref{Proca}) into the equations
(\ref{Dir-1a})-(\ref{Dir-1c}) we obtain the spin 1 Proca equations
\begin{equation} \label{Pr1}
P^\mu A_\mu{}^{ij}=0\, ,\quad P_\mu A_\nu{}^{ij} -P_\nu A_\mu{}^{ij} = F_{\mu\nu}{}^{ij}\, , \quad
P^\mu F_{\mu\nu}{}^{ij}-m^2 A_\nu{}^{ij} =0\, ,
\end{equation}
as well as the identity $P_{[\,\mu} F_{\nu\lambda]}{}^{ij} =0\, .$

We obtained three complex fields (internal $SU(2)$-triplet) with
spin $s~=~1$. On the function (\ref{sol-1}) we impose  the
reality condition $\widetilde\Phi=\overline{\widetilde\Phi}$ which
gives $
\phi^{\dot\alpha\dot\beta}{}_{ij}=\bar\phi^{\dot\alpha\dot\beta}{}_{ij}=
\overline{(\phi^{\alpha\beta}{}^{ij})}\, , \quad \phi^{\alpha
\dot\beta}{}^{i}_{j}=\overline{(\phi^{\beta\dot\alpha}{}^{j}_{i})}
\, .$ The second set of relations in (\ref{Pr1}) can be written down as the following
$SU(2)$-Majorana reality conditions $\psi^{ij}_{cd}(C\gamma_5)_{ca}
(C\gamma_5)_{db}=\epsilon^{ik}\epsilon^{jl}\overline{\psi}_{klab}$
and the fields (\ref{Proca}) satisfy the reality conditions $\overline{(A_\mu{}^{ij})}=A_\mu{}_{ij}\, ,\quad
\overline{(F_{\mu\nu}{}^{ij})}=F_{\mu\nu}{}_{ij}\, .$ These relations define three real vector fields and
the corresponding three real field strengths.

\section{Conclusions}\label{Sect.Conclusions}

We have described a classical and first-quantized model of massive relativistic particles
with spin based on a hybrid geometry of phase space, with primary spacetime coordinates
$x_\mu$ and composite four-momenta $P_\mu$ expressed in terms of fundamental spinorial
variables. We would like to point out that a model for massive particles with spin in an
enlarged spacetime derived from two-twistor geometry, with primary both spacetime
coordinates and four-momenta $P_\mu$, has been recently described in
\cite{BeAzLuMi04}-\cite{AzFrLuMi}. The difference with our approach here consists in the
choice of the primary geometric variables which in \cite{FeZi01}-\cite{Fed95} contains,
besides two-twistor degrees of freedom in mixed twistorial-spacetime formulation, a
primary even Weyl spinor \cite{FeZi95b}.\footnote{It is worth mentioning that the fixing of spin in \cite{FeZi95b}
leads to violation of even ``supersymmetry''. The models with a bosonic counterpart of
supersymmetry describe higher spin particle \cite{FeLu06,FeIv06} and higher spin
superparticle \cite{FeIvLu06}.} In this work all the degrees of freedom describing massive
particles with spin and internal charge are derived entirely from the two-twistor geometry.

In order to quantize the classical system we have introduced a complete set of commuting
observables, which determine the generalized coordinates of the wavefunction. In our case
the set of commuting generalized coordinates does not contain all the spacetime
coordinates, because in our geometric framework they do not commute. As a result, only the
Lorentz-invariant projection $m^2\widetilde{x}_0=x_\mu P^\mu$ can be included into the
quantum-mechanical commuting coordinates. In such a way we are allowed to use the plane waves
$e^{ix_\mu P^\mu}$ as describing the spacetime dependence of the wavefunction. We conclude,
therefore, that although in our framework the spacetime coordinates of spinning massive
particles are non-commutative, we are able to obtain the standard plane wave solutions.

\subsubsection*{Acknowledgements}

Two of us (S.F. and C.M.E.) would like to thank the organizers of the {\it 22nd Max Born Symposium}
for its very pleasant atmosphere. The work of S.F. was supported in part by the RFBR grant
06-02-16684, the grant INTAS-05-7928 and the grants from Bogoliubov-Infeld and
Heisenberg-Landau programs. Two of the authors (A.F.
and J.L) would like to acknowledge the support by KBN grant 1
P03B 01828. The work of C.M.E. has been supported  by research grants from
the Spanish Ministerio de Educaci\'{o}n y Ciencia (FIS2005-02761 and EU FEDER funds)
, the Generalitat Valenciana and by the EU network MRTN-CT-2004-005104 ('Forces Universe').
One of us (C.M.E.) wishes to thank Prof. Siegel and the members of the YITP of Stony Brook for the
kind hospitality extended to him. Discussions with Jos\'{e} de Azc\'{a}rraga are also acknowledged.

\end{document}